\documentclass[superscriptaddress,aps,pra,twocolumn,showpacs,nofootinbib,longbibliography]{revtex4-1}
\usepackage{amsmath,amsthm}
\usepackage{algorithm}
\usepackage{algpseudocode,caption}
\usepackage{braket}
\usepackage{array}
\usepackage{amsmath}
\usepackage{amsfonts}
\usepackage{amssymb}\usepackage[T1]{fontenc}
\usepackage{latexsym}
\usepackage{amssymb}
\usepackage[colorlinks=true,citecolor=blue,urlcolor=blue]{hyperref}
\usepackage{color}
\usepackage{graphics,epstopdf}
\usepackage{soul}
\usepackage[demo]{graphicx}
\usepackage{subcaption}
\usepackage{lipsum}
\usepackage{float}
\usepackage{adjustbox}
\usepackage[normalem]{ulem}
\newlength\myindent
\newcommand{\be}{\begin{equation}}
\newcommand{\ee}{\end{equation}}
\newcommand{\ba}{\begin{eqnarray}}
\newcommand{\ea}{\end{eqnarray}}

\begin{document}
	
\title{Quantum entropy expansion using n-qubit permutation matrices in Galois field}

\author{Avval Amil}
\email{avval2801@gmail.com}
\affiliation{Department of Computer Science and Engineering, I.I.T. Delhi, Hauz Khas, New Delhi - 110016, India}

\author{Shashank Gupta}
\email{shashank@qnulabs.com}
\affiliation{QuNu Labs Pvt. Ltd., M. G. Road, Bengaluru, Karnataka 560025, India}

\date{\today}

\begin{abstract}

Random numbers are critical
for any cryptographic application. However, the data that is flowing through the internet is not secure because of entropy deprived pseudo random number generators and unencrypted IoTs. In this work, we address the issue of lesser entropy of several data formats. Specifically, we use the large information space associated with the n-qubit permutation matrices to expand the entropy of any data without increasing the size of the data. We take English text with the entropy in the range 4 - 5 bits per byte. We manipulate the data using a set of n-qubit (n $\leq$ 10) permutation matrices and observe the expansion of the entropy in the manipulated data (to more than 7.9 bits per byte). We also observe similar behaviour with other data formats like image, audio etc. (n $\leq$ 15).

\end{abstract}

\maketitle
\section{Introduction}
The significance of random numbers in human civilisation is unparalleled. Independent and identically distributed random numbers are building blocks for cryptographic, chemical, biological, industrial, logistic, and financial applications \cite{Gisin2002}. Entropy starvation is the fundamental back door for various cryptographic attacks. Depending upon the characteristic of the source, randomness can be epistemic (apparent) or ontic (intrinsic) \cite{Penrose37}. These two notions of randomness give rise to different categories of random number generators.

Pseudo random number generators (PRNGs) generate random numbers using some initial seed and mathematical algorithm and hence are deterministic \cite{Gisin2002, Luca20, George03, Vigna14}. On the other hand, random numbers derived from some physical processes are called true random number generators (TRNGs) \cite{Gisin2002, James20}. Within TRNGs, there is a class of random number generators that generate, expand and amplify the randomness using some quantum phenomenon called quantum random number generators (QRNGs). At present, QRNGs are gaining significant attention because of the intrinsic random behaviour of quantum mechanics \cite{Ma16, Tobias21, Nie16}. A class of QRNGs utilizes nonclassical resources like non-locality \cite{Gisin2002}, contextuality \cite{Gupta22}, quantum steering \cite{Gupta21} to generate self-testable randomness. The underlying process in a QRNG is simple and therefore preferable over the complex chaotic classical phenomenon in other TRNGs. Commercial QRNGs are available in the market from various vendors like ID Quantique \cite{ID17}, QNu Labs \cite{Qnu20}, Quintessence Labs \cite{Quintessence20} etc. However, even QRNGs face manufacturing bias, that is corrected using extractors or privacy amplification \cite{Dodis13, Darren17}. 

QRNGs haven't been able to replace the PRNGs because of the higher cost and integration difficulties in diverse scenarios. This creates a possibility of cyber attacks in the traditional applications because of entropy starvation and unavailability of encryption in some resource constrained IoT devices \cite{Amalia17, Randy21}. In this work, we attempt to address this question. In particular, we present an entropy expansion algorithm that increases the entropy of any filetype without modifying its size. Hence, our algorithm can be deployed before sending the files for a secure communication. Thus, it addresses the problem of entropy starvation in a cost effective way for resource constrained IoT devices and other tradition cryptographic applications.

Our algorithm is based on the large information space associated with the n-qubit permutation matrices ($2^{n}!$) \cite{Bona04}. On the other hand, an n-qubit system has a $2^n$ information state represented  by  a  Galois  field  GF($2^n$). The  entire set of permutation matrices form the symmetric group $S_{2^n}$, with a total of $(2^{n}!)$ unique permutations. An arbitrary permutation matrix can be physically implemented using an algorithm proposed by Shende et al. in 2003 \cite{Shende03}, using elementary quantum gates like CNOT and TOFFOLI, and can also be mathematically expressed  using classical computing systems. An n-bit permutation space can be considered as an entropy expansion from  the classical  Boolean  information space or Galois field GF($2^n$)  to  quantum  permutation  space,  or $S_{2^n}$ \cite{Kuang21}. This huge entropy from the quantum permutation space has the properties of Shannon perfect secrecy which lays a foundation for quantum safe cryptography \cite{Randy20}.

In the present work, we investigate the expansion of entropy of several data types using n-qubit permutation matrices. In particular, we analyse the efficacy of a set of permutation matrices in expanding the entropy of general English text, images etc.
In each scenario, we generate a set of permutation matrices of different dimensions (N = 16, 128, 1024 etc.) and analyse the results for the expansion of entropy. 
We demonstrate that our algorithm works in any scenario without changing the size of the data file. The entropy estimation and other randomness parameter are determined using ENT randomness test \cite{ENT}. There are other popular randomness test suites like NIST SP 800-22 \cite{NIST}, Diehard \cite{diehard}, test-U01 \cite{test01}, etc. We stick to ENT for the purpose of randomness parameter estimation.  

The paper is organised in the following way. In Sec.(\ref{Prelim}), we recapitulate the set of n-qubit permutation matrices and the generation of an n-qubit permutation matrices using QRNG data from Qosmos. In Sec.(\ref{DI}), we discuss the general entropy expansion algorithm using the generated permutation matrices. In Sec. (\ref{data}), we present our entropy expansion results in three different scenarios, 1. Random English text. 2. Detailed (higher number of colours, larger variation in colours) as well as simple images. 3. Audio files.  Finally, we
conclude in Sec.(\ref{Conclusion}) with some future offshoots of the present analysis.

\section{Preliminaries}
\label{Prelim}
\textit{Notation:} In this paper we will be using the following notation. In an n-qubit quantum system, there are $N = 2^n$ computational basis vectors and thus any state of an n-qubit system can be represented by a vector of size $N = 2^n$. Also, quantum mechanical operators (e.g. permutation operators which we will be focusing on in this paper) can be represented by square matrices of size $N\times N = 2^n \times 2^n$ . These operators operate on the system's state, transforming it to another state. Mathematically, this is equivalent to a vector being multiplied by a matrix.\\ 
\textit{Permutation matrices:} A permutation matrix is a square matrix with the property that each row and each column of the matrix contains exactly one non-zero entry which is equal to 1. Note that the identity matrix is a permutation matrix. In fact, any permutation matrix can be generated from the identity matrix by simply permuting its rows (or columns). For an n-qubit system, the identity matrix has $ N= 2^n$ rows, and thus the number of permutation matrices possible is $(2^n)!$  which is the number of ways to permute the $N$ rows. This large size of the information space represented by permutation matrices is key in the entropy expansion of data  \cite{Kuang21}.\\
\textit{Generalized permutation matrices:} The set of generalized permutation matrices is a superset of the set of permutation matrices. Like permutation matrices, these also have exactly one non-zero entry in each row and column, but the difference is that this non-zero entry can be any number and not just 1.\\
\textit{Examples of Permutation Matrices:} Some permutation matrices for a 2-qubit system ($N=2^2 = 4)$ can be -
\begin{center}
    $\begin{bmatrix}
        1 & 0 & 0 & 0\\
        0 & 0 & 1 & 0\\
        0 & 0 & 0 & 1\\
        0 & 1 & 0 & 0
    \end{bmatrix}
    \begin{bmatrix}
        0 & 0 & 1 & 0\\
        0 & 0 & 0 & 1\\
        1 & 0 & 0 & 0\\
        0 & 1 & 0 & 0
    \end{bmatrix}
    \begin{bmatrix}
        1 & 0 & 0 & 0\\
        0 & 0 & 1 & 0\\
        0 & 1 & 0 & 0\\
        0 & 0 & 0 & 1
    \end{bmatrix}$
\end{center}
\textit{Examples of Generalized Permutation Matrices:} Some generalized permutation matrices can be -
\begin{center}
    $\begin{bmatrix}
        0 & \sqrt{2} & 0 & 0\\
        0 & 0 & 3 & 0\\
        \pi & 0 & 0 & 0\\
        0 & 0 & 0 & 1
    \end{bmatrix}
    \begin{bmatrix}
        1 & 0 & 0 & 0\\
        0 & 2 & 0 & 0\\
        0 & 0 & 3 & 0\\
        0 & 0 & 0 & 4
    \end{bmatrix}
    \begin{bmatrix}
        0 & 0 & 0 &\sqrt{5}\\
        0 & e & 0 & 0\\
        10 & 0 & 0 & 0\\
        0 & 0 & -5 & 0
    \end{bmatrix}$
\end{center}

\subsection{Generation of n-qubit Permutation Matrices}
The entropy expansion algorithm (discussed subsequently) requires permutation matrices and for their generation, we have used the Fisher-Yates shuffle algorithm. A description of the same is given now in the form of pseudocode. Note that all matrices and arrays are indexed from $1$ to $N$ and not $0$ to $(N-1)$ (for matrix size $N \times N$).\\ \\
\textbf{Permutation Matrix Generation Using Fisher-Yates Shuffle Algorithm:}\\
\begin{algorithmic}[1]
\Require RandomInt(1,$N$) returns a random integer between 1 and $N$ (inclusive) when called; Swap($a$,$b$) swaps the values of $a$ and $b$
\Ensure $P$ is a random permutation matrix
\State $i \leftarrow 1$
\While{$i \leq N$}
\State $K[i] \leftarrow$ RandomInt(1,$N$)
  \State $S[i] \leftarrow i$
  \State $j \leftarrow 1$
    \While{$j \leq N$}
        \State $P[i][j] \leftarrow 0$
        \State $j \leftarrow (j+1)$
    \EndWhile
    \State $i \leftarrow (i+1)$
\EndWhile
\State $i\leftarrow N$
\While{$i>0$}
    \State $p \leftarrow K[i]$
    \State $Swap(S[p],S[i])$
    \State $i \leftarrow (i-1)$
\EndWhile
\State $i \leftarrow 1$
\While{$i\leq N$}
    \State $P[i][S[i]] \leftarrow 1$
    \State $i \leftarrow (i+1)$
\EndWhile
\end{algorithmic}
For the above algorithm to work, we need a source of random numbers (to use in RandomInt(1,$N$)). These random numbers can be created from any random data such as those generated by QRNGs (we have used Qosmos (QNu Labs' Entropy-as-a-Service) for the same). The algorithm described above creates an $N \times N$ matrix, but in our implementation we have used the fact that permutation matrices are sparse in nature and used a sparse representation to work with them. The underlying idea is the same however.
\subsection{Crux of Entropy Expansion}
For $n$ classical bits which are transformed by Boolean algebra, the information space is of size $2^n$. If we fix a type of Boolean operation (e.g. AND, OR, XOR) to be done on the $n$ bits there are $2^n$ possible mappings of inputs to outputs. This is because there are $2^n$ bit-strings possible with which we can perform the operation chosen on the input. \\
On the other hand, for $n$ qubits which are transformed by linear algebra, if we fix the type of linear transformation to be a permutation, then there are $(2^n)!$ possible mappings of inputs to outputs due to the $(2^n)!$ different permutation matrices available. These permutation matrices form a permutation space of $(2^n)!$ dimensions with a Shannon entropy of log$_2((2^n)!) \in \Theta (n2^n)$ (Big Theta $\Theta$ asymptotic notation - this result can be proved by seeing that for a large positive number $m$, log$(m!) = \sum_{k=1}^{m}$log$(k)\leq m$log$(m)$ and also log$(m!) = \sum_{k=1}^{m}$log$(k) \geq \sum_{k=m/2}^{m}$log$(k) \geq \frac{m}{2}$log$(\frac{m}{2})$ and replacing $m$ with $2^n$). The use of permutation matrices can thus give an entropy expansion from the classical Boolean information space to the quantum permutation space \cite{Kuang21}.\\
Using quantum hardware, entropy expansion can be implemented using quantum gates (any permutation gate can be physically implemented using NOT, CNOT and TOFFOLI gates\cite{Shende03}). But as the permutation gates belong to a special subset of quantum gates which have classical behaviour, the entropy expansion algorithm can be implemented on classical systems and the increase in entropy can be verified.

\section{Entropy expansion algorithm}\label{DI}
An important point to note about the entropy expansion algorithm is that it needs data in the form of a binary string - a string of only zeroes and ones (e.g. `00110110'). What this data represents depends on the format we are working with. For a text file, this data can be the ASCII values of each character concatenated in the order they appear in the text; for an image, it can be the RGB values of the pixels in binary concatenated according to their position in the image and so on. \\
The algorithm takes three inputs - the bit-string just described above, the size of the permutation matrix we need to work with (for an n-qubit system this size is $2^n$) and the number of permutation matrices. The number of permutation matrices possible for an n-qubit system is $(2^n)!$, but we obviously can't work with a huge number of matrices like 32! or 64!, and hence the number of matrices we will be working with is taken as an input. \\
After taking the inputs, the first step in the algorithm is to create a subset of the permutation matrices by repeated application of the Fisher-Yates shuffle algorithm. The number of matrices and their sizes in this step were defined at the input.\\
Next, the bit-string is broken down into contiguous chunks. If the matrices are of size $N\times N$, then the size of each chunk will be $N$. Each chunk here will behave as an $N$-dimensional binary vector.\\
Now for each chunk we select a permutation matrix out of the subset we had generated (according to some procedure, this can vary - in our case when working with $m$ permutation matrices, we simply generated $log_2(m)$ random bits using a QRNG and selected the matrix according to the number specified by those bits), and multiply that matrix with the binary vector represented by that chunk. This will be repeated for each chunk, and the resultant permuted vectors are concatenated to give a permuted bit-string. The permuted bit-string can be used to recreate the format being worked with (text, image, audio) and the result can be compared with the original. It can be seen that the permuted bit-string is of the same length as the input bit-string if the length is a multiple of the size of the matrices, if this is not the case then we can either drop the last chunk or pad the original bit-string suitably. An issue that may arise with padding (e.g. if we pad with all zeroes) is that the padding may decrease the randomness (entropy) of the input and hence the output too. 

\begin{figure}[h!]
    \centering
    \resizebox{8.4cm}{12cm}{\includegraphics{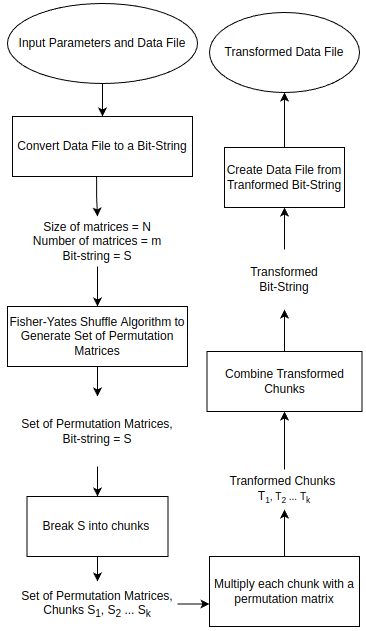}}
    \caption{Flowchart Explaining the Entropy Expansion Algorithm}
    \label{expansionalgo}
\end{figure}

\subsection{Analysis and Performance of Algorithm}
If the number and size of matrices are constant, then we claim that this algorithm can be executed in $\mathcal{O}(l)$, where $l$ is the length of the bit-string. Let the size of the matrices be $N$. The number of chunks will then be $\frac{l}{N}$. If we use traditional matrix multiplication, then multiplying an $N\times N$ matrix with an $N\times 1$ vector is $\mathcal{O}(N^2)$, and as the number of multiplications is $\frac{l}{N}$, the total time taken will be $\mathcal{O}(N^2*\frac{l}{N}) = \mathcal{O}(Nl)$ which is $\mathcal{O}(l)$ if the size of matrices is constant. \\
A better implementation is possible for the algorithm if we use one simple observation - multiplying a binary vector by a permutation matrix only shuffles the positions of each entry in the vector. Using this observation, the matrix multiplication can be reduced to $\mathcal{O}(N)$ and the total time taken will be $\mathcal{O}(N*\frac{l}{N}) = \mathcal{O}(l)$.\\
If we vary the number of matrices, keeping the size and bit-string constant, then the time taken for the generation of the matrices increases, and the selection process also needs more data so the overall time increases. \\
If we vary the size of the matrices, keeping the number and bit-string constant, then we have some interesting observations. The generation of the matrices obviously takes longer, but as the number of chunks decreases, there is a trade-off in the time taken. It has been observed that the algorithm takes less time to execute when working with $256\times 256$ matrices as compared to $32\times 32$
matrices, and the time is even less for $1024\times 1024$ matrices. This will be the case till the matrix size is sufficiently large and the time taken for the creation of the matrix starts to dominate the time taken for the manipulation of the bit-string. This behaviour is favorable because (as shown in the next section) the degree of entropy expansion improves on using larger matrix sizes, and this reduced time of execution is an added benefit for practical use.

 \section{Results}\label{data}
 
\subsection{English Text}
We first tested the algorithm on various small files containing English text. Randomness testing was done on both the original text files and the transformed text files. For this purpose, the ENT randomness test was used\cite{ENT}. All modifications listed here were performed using 16 matrices of size $256 \times 256$ (Table \ref{table1} and Table \ref{table2}). The bit-string for English texts was constructed by considering the ASCII values of each character and then concatenating them all.\\
The testing gave results for 5 parameters related to randomness of the data - entropy, chi-square distribution, arithmetic mean, Monte Carlo value of $\pi$ and the serial correlation coefficient. In all the cases tested, an improvement in the values of all the parameters was observed (deviation from their ideal values decreased). The results are compiled in the tables below. After this, we tested the same for a bigger text file of size 10MB, and the results obtained are also presented. In this case all the parameters improved, with the exception of the serial correlation coefficient.\\
Overall, for the first 5 smaller files (varying sizes), the entropy showed an increase of $75.08\%$, $78.09\%$, $82.55\%$, $77.99\%$ and $69.69\%$ respectively, while the larger (10MB) file showed an increase of $69.18\%$. \\The absolute deviations from the ideal value of the Chi-Square distribution showed a decrease of $99.40\%$, $99.33\%$, $99.41\%$, $99.64\%$, $99.28\%$ and $99.25\%$ for the six files respectively. \\The absolute deviations from the ideal value of the arithmetic mean showed a decrease of $74.10\%$, $64.74\%$, $54.30\%$, $65.52\%$, $56.83\%$ and $38.94\%$ for the six files respectively. 
\begin{widetext}
\begin{center}
\begin{table}[h!]
\centering
    \begin{tabular}{| m{2cm} | m{2cm} | m{2cm} | m{2cm} | m{2cm} | m{2cm} | m{2cm} | m{2cm} |}
        \hline
        Parameter & Input File 1& Output File 1& Input File 2& Output File 2& Input File 3& Output File 3&Ideal Value\\
        \hline \hline
        Entropy & 4.443597 & 7.779930 & 4.415748&7.863862 &4.285194& 7.822699&8.000000\\
        \hline
        Chi-Square Distribution & 20511.63 & 376.40&58516.92& 644.08& 34801.26&458.40&256.00 \\
        \hline
        Arithmetic Mean & 92.2588 & 118.3742&92.4348& 115.1365&91.1942& 110.9078&127.5\\
        \hline
        Monte Carlo value of Pi & 4.000000000 & 3.305164319 &3.992970123&3.299115044&4.000000000& 3.550000000&3.141592653\\
        \hline
        Serial Correlation Coefficient & -0.083482 & 0.014627 &-0.089758&0.057929& -0.082719&0.026761&0.0\\
        \hline
    \end{tabular}
    \caption{Results for 1306 byte (File 1), 3418 byte (File 2) and 1941 byte (File 3) English text files}
    \label{table1}
\end{table}
\end{center}
\begin{center}
\begin{table}[h!]
\centering
    \begin{tabular}{| m{2cm} | m{2cm} | m{2cm} | m{2cm} | m{2cm} | m{2cm} | m{2cm} | m{2cm} |}
        \hline
        Parameter & Input File 4& Output File 4& Input File 5& Output File 5& Input File 6& Output File 6&Ideal Value\\
        \hline \hline
        Entropy & 4.416907 & 7.861833 & 4.618982&7.837981 &4.700438& 7.952263&8.000000\\
        \hline
        Chi-Square Distribution & 30440.22 & 364.65&32108.75& 484.47& 92758880.84&696321.91&256.00 \\
        \hline
        Arithmetic Mean & 93.6210 & 115.8191&95.2057& 113.5579&109.5021& 138.4849&127.5\\
        \hline
        Monte Carlo value of Pi & 4.000000000 & 3.381818182 &3.978021978&3.447513812&4.000000000& 2.910716595&3.141592653\\
        \hline
        Serial Correlation Coefficient & -0.024084 & -0.011061 &-0.056893&0.042615& -0.000179&0.005188&0.0\\
        \hline
    \end{tabular}
    \caption{Results for 1987 byte (File 4), 2188 byte (File 5) and 10MB (File 6) English text files}
    \label{table2}
\end{table}
\end{center}
\end{widetext}

\subsection{Effect of Size and Number of Matrices on Entropy Expansion}
\begin{figure*}
    \begin{subfigure}{8.5cm}
    \centering\includegraphics[width=8.5cm]{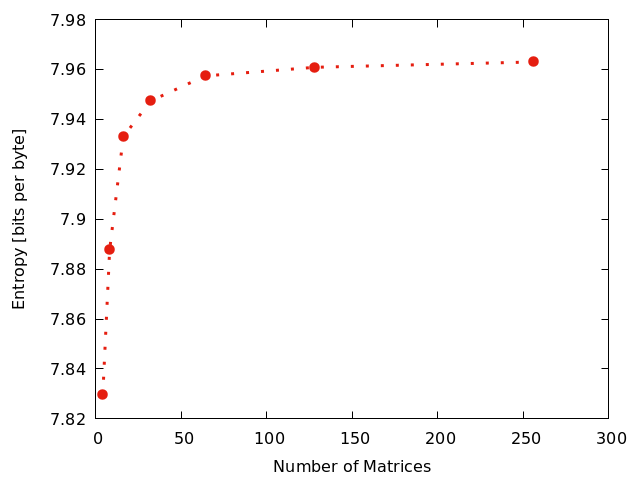}
    \caption{Entropy variation with no. of matrices in the set by keeping the Size of each matrix as ($128 \times 128$) constant.}
    \label{graph_a}
    \end{subfigure}%
    \begin{subfigure}{8.5cm}
    \centering\includegraphics[width=8.5cm]{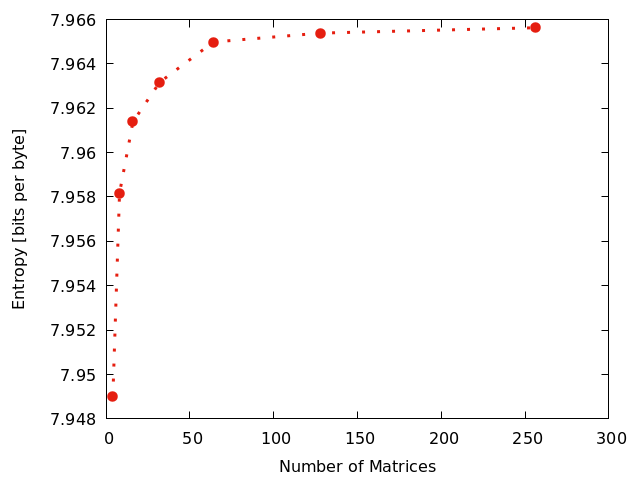}
    \caption{Entropy variation with no. of matrices in the set by keeping the size of each matrix as ($1024 \times 1024$) constant.}
    \label{graph_b}
    \end{subfigure}
    \begin{subfigure}{8.5cm}
    \centering\includegraphics[width=8.5cm]{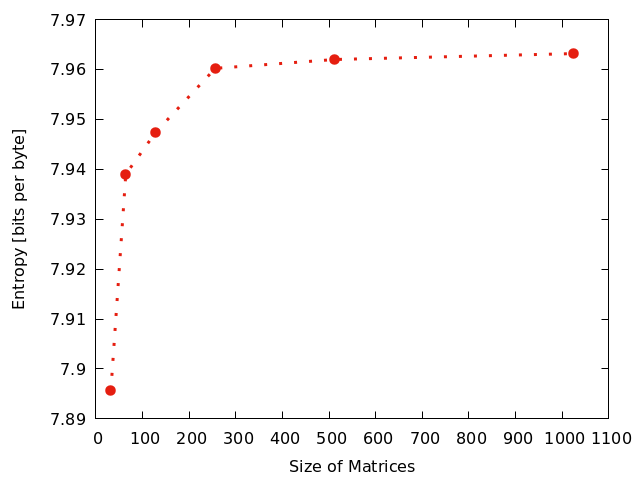}
    \caption{Entropy variation with size of matrices keeping number of matrices in the set as (32) constant.}
    \label{graph_c}
    \end{subfigure}
    \begin{subfigure}{8.5cm}
    \centering\includegraphics[width=8.5cm]{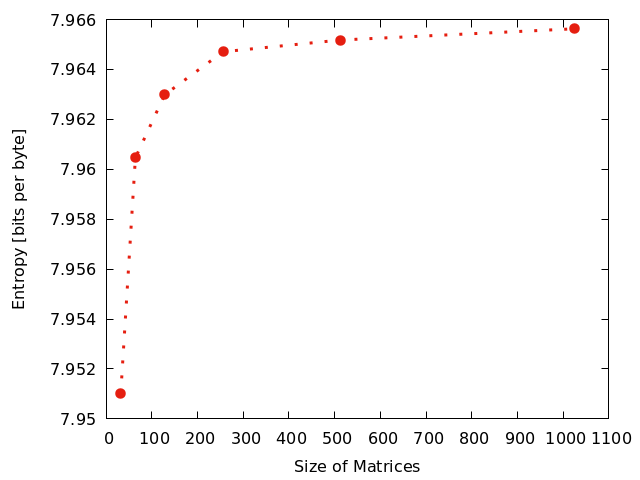}
    \caption{Entropy variation with size of matrices keeping number of matrices in the set as (256) constant.}
    \label{graph_d}
    \end{subfigure}%
\caption{Entropy variation (on 10MB English text file) with size and number of the matrices in the set.}
\label{graphs}
\end{figure*}

All of the above testing was done using 16 matrices of size $256\times 256$. To understand the effect of using different matrix sizes and number of matrices on entropy expansion, we varied these two quantities for a single file - the 10MB text taken above. We observed that as the size of the matrices increases, the entropy of the modified file also increases and seems to approach an asymptote. It was also observed that as the number of matrices used increase, entropy shows a similar behaviour and increases while seeming to approach an asymptote (Fig. \ref{graphs}).\\
Thus, for maximum increase in the entropy of the file, the ideal combination seems to use the maximum number and maximum size of matrices possible.\\
The following graphs (Fig. \ref{graphs}) show the behaviour of entropy expansion in four cases -
\begin{itemize}
    \item Number of matrices in the set is varied keeping size as ($128 \times 128$) constant (fig. \ref{graph_a}). 
    \item Number of matrices in the set is varied  keeping size as ($1024 \times 1024$) constant (fig. \ref{graph_b}).
    \item Size of matrices is varied keeping number of the matrices in the set as (32) constant (fig. \ref{graph_c}).
    \item Size of matrices is varied keeping number of matrices in the set as (256) constant (fig. \ref{graph_d}).
\end{itemize}

Similar plots were also made for other matrix sizes and numbers, and the same behaviour was noted in all of them. It seems that the size of matrices used plays a greater role in entropy expansion than the number of matrices used, as steeper increase is noticed when varying the size of matrices as compared to the number of matrices. This may be due to the fact that there is an exponential increase in the size of the permutation space when increasing the matrix sizes, while only a linear increase with the number of matrices.\\
The most interesting observation here is the presence of an asymptote, for which there could be many explanations. Of course the entropy of all files has an upper bound of 8 bits per byte, and it could simply be the case that the entropy is increasing towards 8 bits per byte. But other factors could also be at play here - for example, the size and contents of the file.
\subsection{Effect of Repeated Application of Entropy Expansion}
In all of the cases observed, we saw that the entropy expansion algorithm did increase the entropy without fail. Next, we observed the effect of repeated application of the entropy expansion algorithm. It was observed that even here, entropy increases with every successive application of the algorithm, but the subsequent increase diminishes after every application. The 10MB English text taken was manipulated using 16 matrices of size $256\times 256$ and the results are tabulated below (Table \ref{repeat}).
\begin{center}
\begin{table}[h!]
\centering
    \begin{tabular}{| m{4cm} | m{4cm} |}
        \hline
        Number of Applications & Entropy\\
        \hline \hline
        0 & 4.700438\\
        \hline
        1 & 7.964177 \\
        \hline
        2 & 7.965788\\
        \hline
        3 & 7.965817 \\
        \hline
        4 & 7.965818\\
        \hline
        5 & 7.965819 \\
        \hline
    \end{tabular}
    \caption{Results for Repeated Application of the Entropy Expansion Algorithm on 10MB File}
    \label{repeat}
\end{table}
\end{center}
The entropy seems to approach a saturation value after 4-5 applications of the algorithm.
\subsection{Images}
\begin{figure*}
    \begin{subfigure}{8cm}
    \centering\includegraphics[width=8cm]{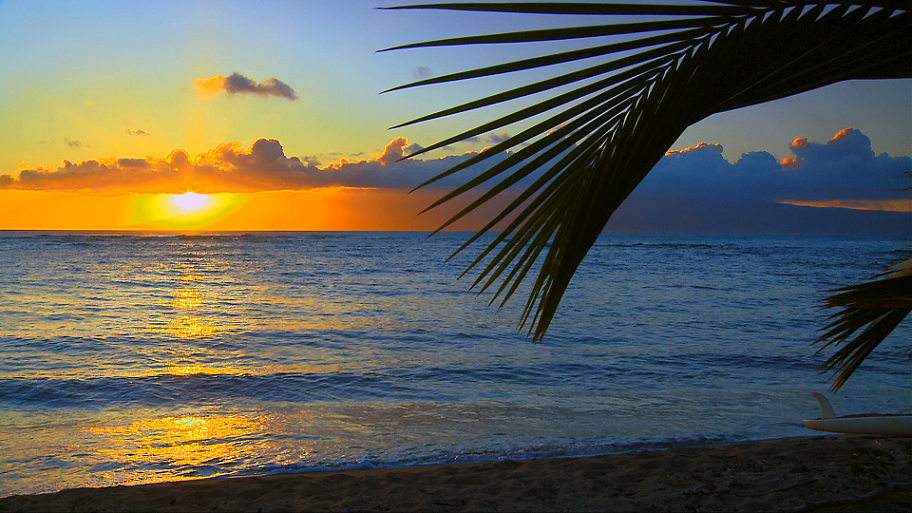}
    \caption{Original Image}
    \end{subfigure}%
    \begin{subfigure}{8cm}
    \centering\includegraphics[width=8cm]{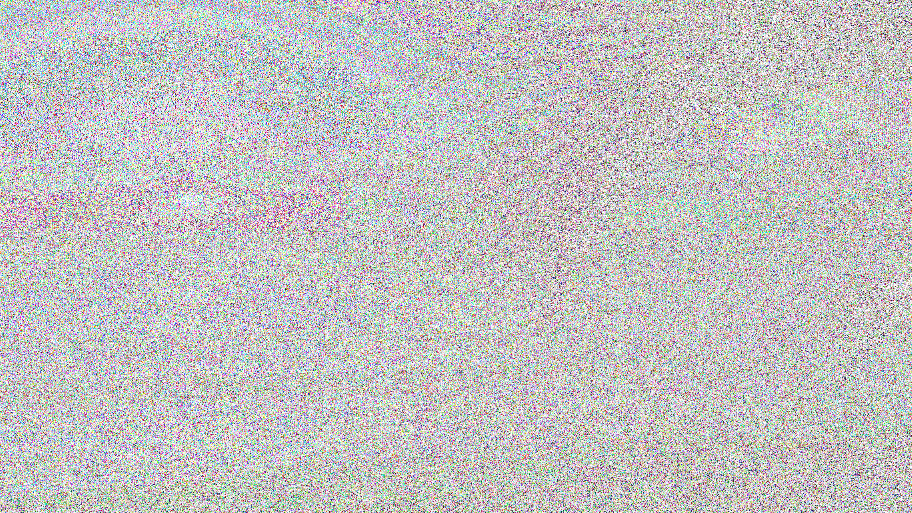}
    \caption{After Entropy Expansion using $64\times 64$ Size Matrices}
    \end{subfigure}
    \begin{subfigure}{8cm}
    \centering\includegraphics[width=8cm]{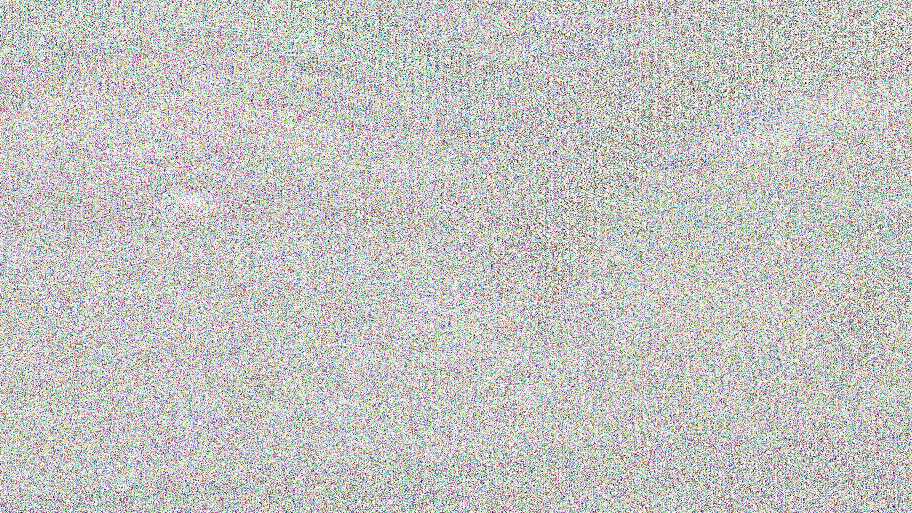}
    \caption{After Entropy Expansion using $256\times 256$ Size Matrices}
    \end{subfigure}
    \begin{subfigure}{8cm}
    \centering\includegraphics[width=8cm]{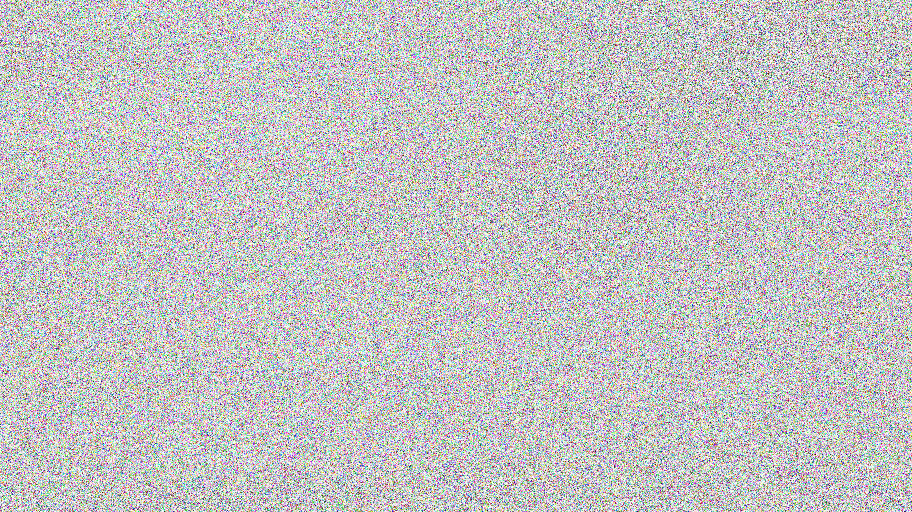}
    \caption{After Entropy Expansion using $8192\times 8192$ Size Matrices}
    \end{subfigure}%
\caption{Effects of Entropy Expansion on an Image}
\label{scene}
\end{figure*}

We also tested the entropy expansion algorithm on images. The main difference here is in the process of creation of the bit-string. As all of the binary of image files is not used to specify how the image looks, we cannot directly create the bit-string by concatenating the binary contents of the file like we did for text files. So here, we created the bit-string by the following process - starting from the top left corner of the image, we took the RGB (in some cases RGBA) values of each pixel, converted them to binary and concatenated all the values. This bit-string could now be transformed by entropy expansion and used to create another image and the two can be compared. The comparison of the visuals also seems to confirm that higher matrix sizes give better improvement in the randomness of data. \\
Now we show an image and the effects of applying entropy expansion to it using different matrix sizes (fig. \ref{scene}).

\begin{figure*}
    \begin{subfigure}{8cm}
    \centering\includegraphics[width=8cm]{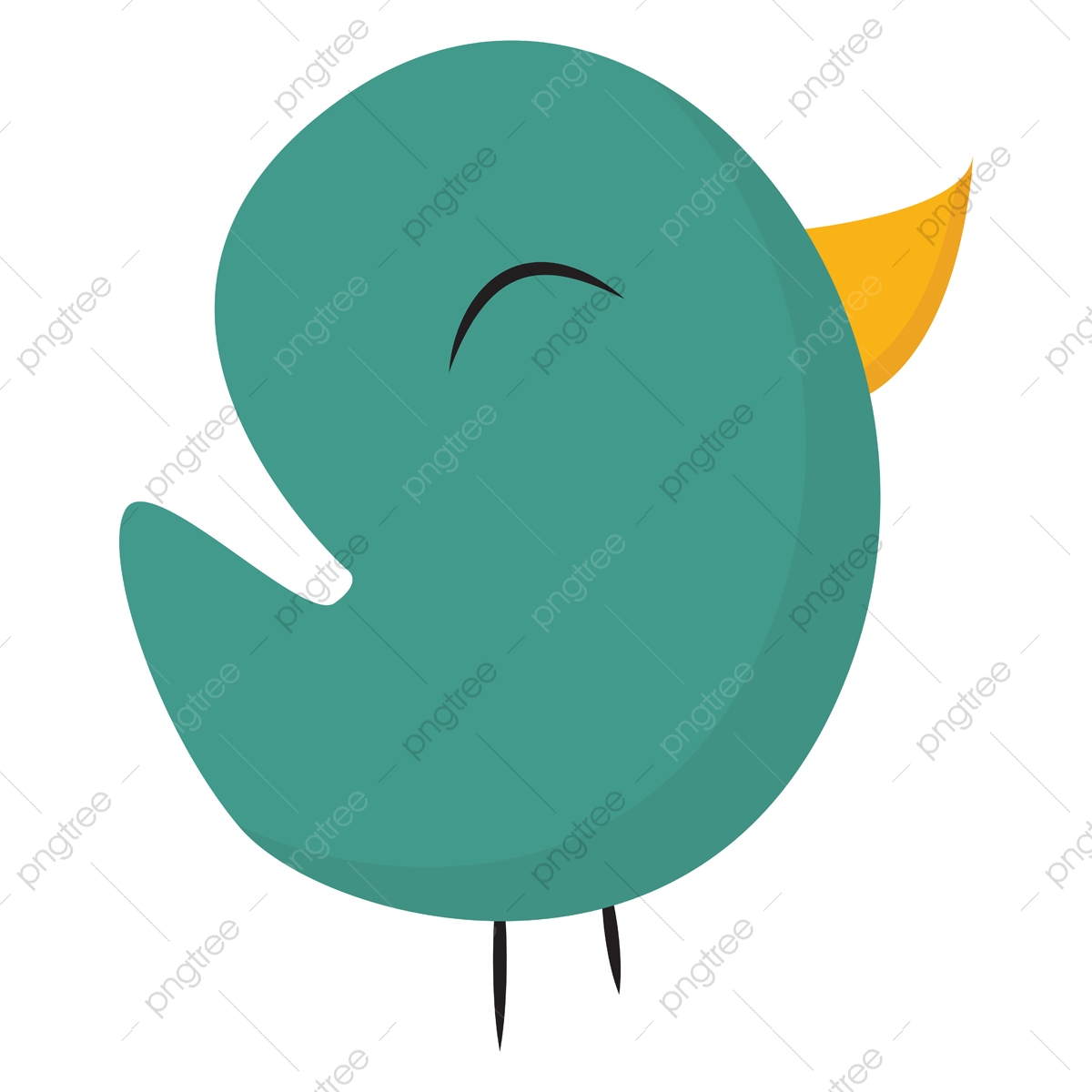}
    \caption{Original Image}
    \end{subfigure}%
    \begin{subfigure}{8cm}
    \centering\includegraphics[width=8cm]{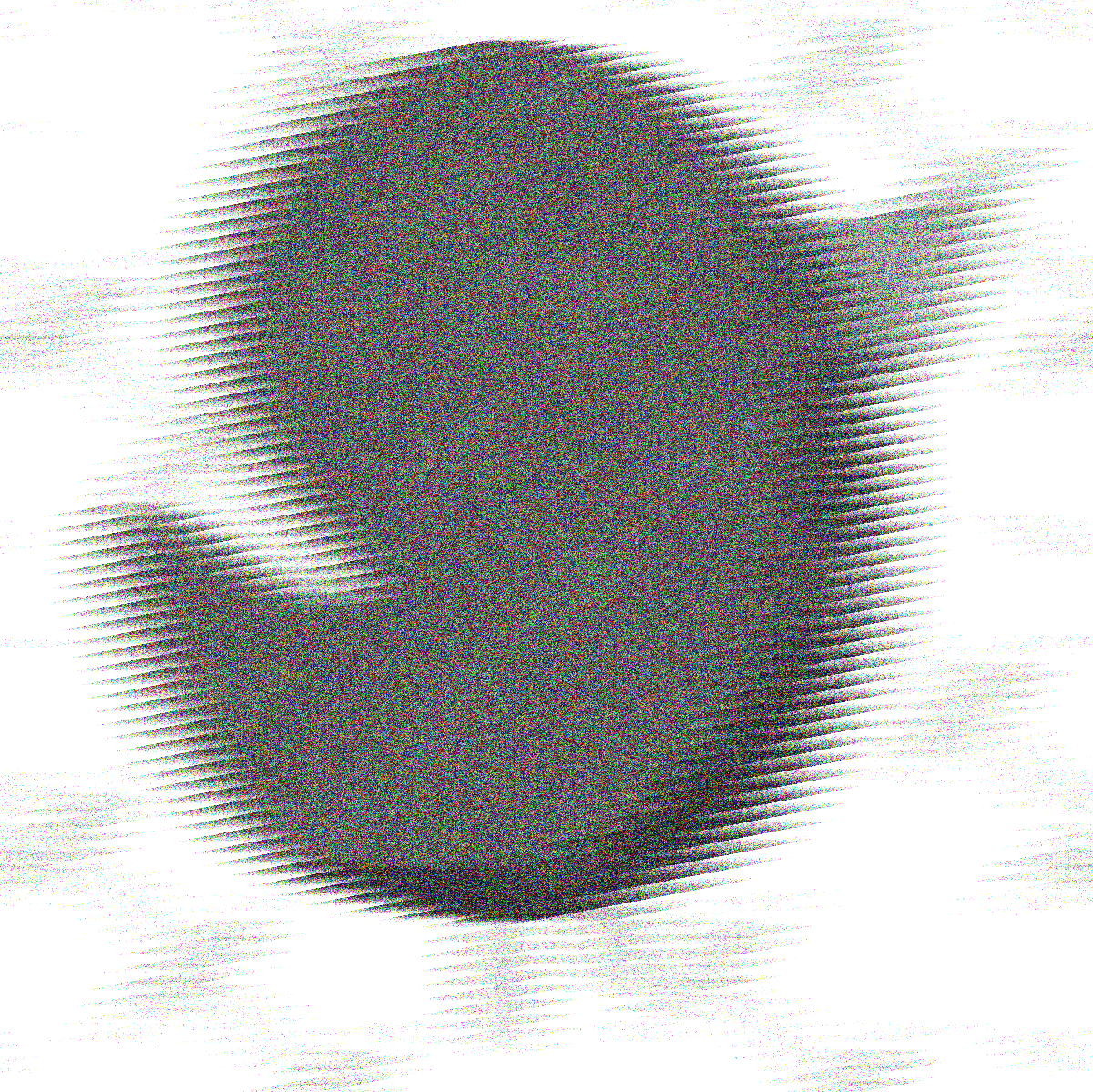}
    \caption{After Entropy Expansion using $2048\times 2048$ Size Matrices}
    \end{subfigure}
    \begin{subfigure}{8cm}
    \centering\includegraphics[width=8cm]{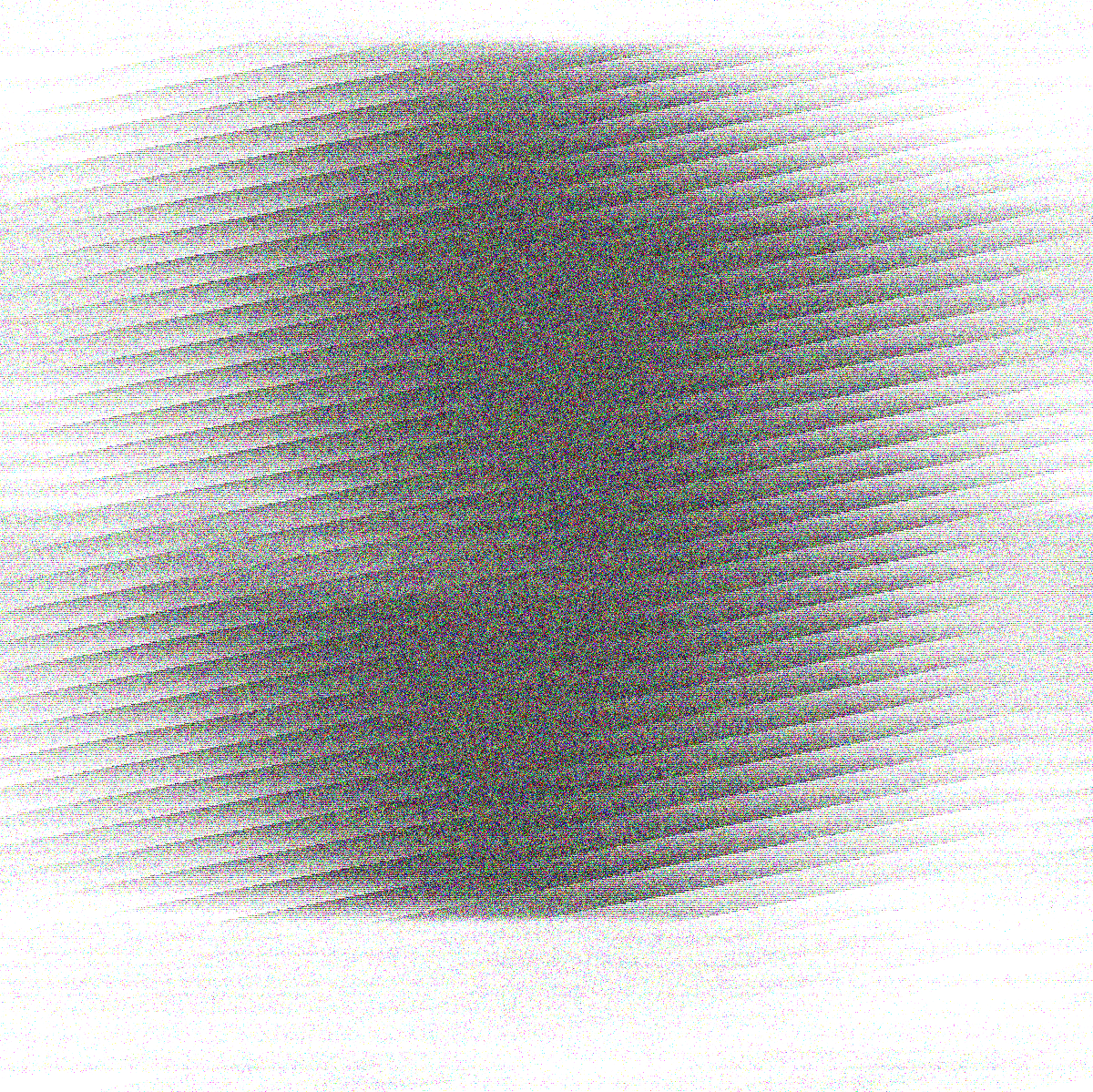}
    \caption{After Entropy Expansion using $8192\times 8192$ Size Matrices}
    \end{subfigure}
    \begin{subfigure}{8cm}
    \centering\includegraphics[width=8cm]{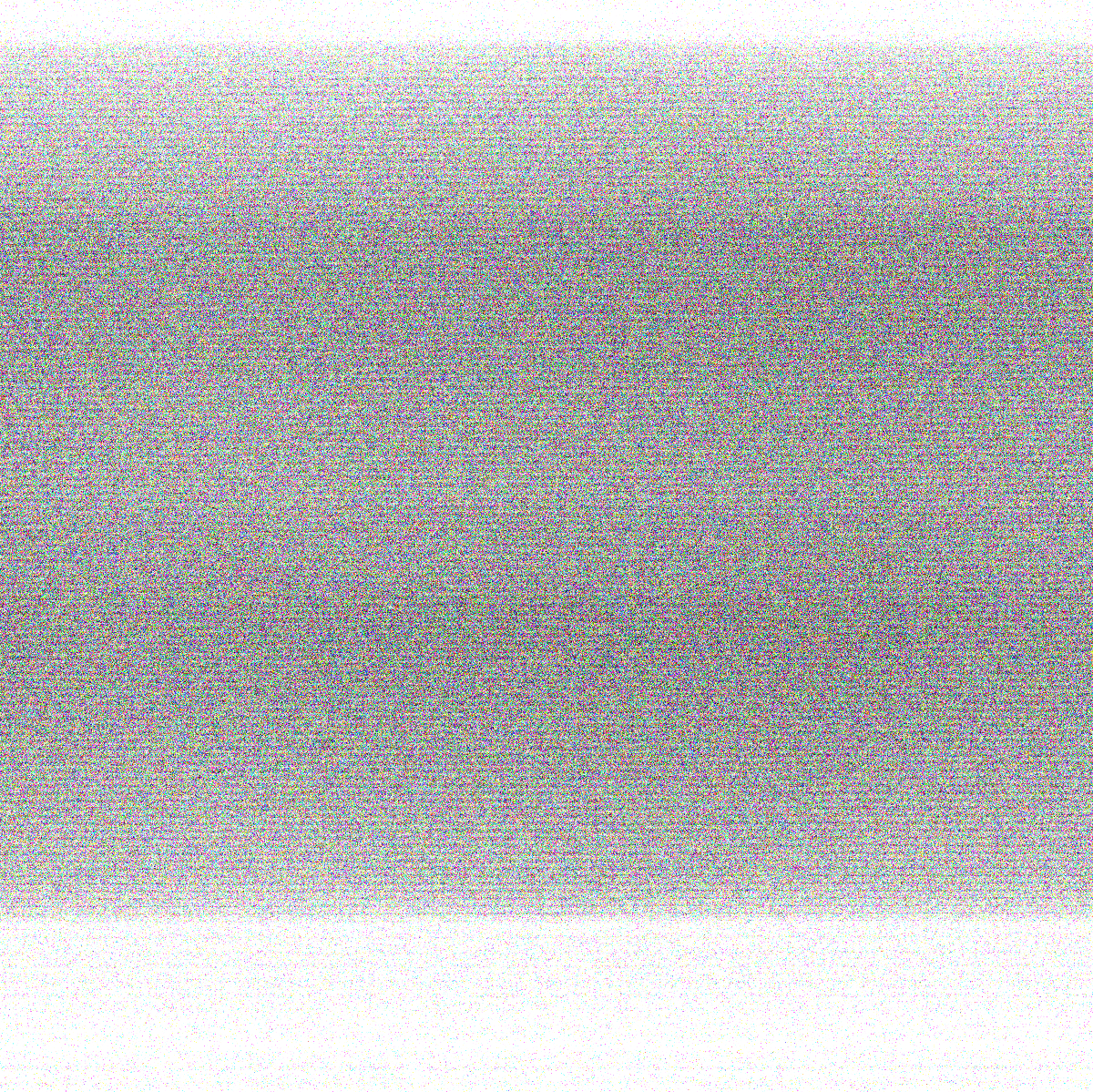}
    \caption{After Entropy Expansion using $32768\times 32768$ Size Matrices}
    \end{subfigure}%
\caption{Effects of Entropy Expansion on a More Uniform Image}
\label{blue_bird}
\end{figure*}

It is visible that using larger matrices distorts the image more and makes it seem more uniformly random. A noteworthy point we observed here was that when working with images which are not that complex originally (more or less uniform with just handful of colours), good randomness is only seen with matrices of sizes near $2^{15}$. This was to be expected as permutation matrices simply shuffle the contents of bytes, and if the bytes are more or less uniform then the shuffling will result in less randomness being created. This can be seen in the following example (fig. \ref{blue_bird}).

\subsection{Audio file}
We can also apply entropy expansion on audio files to further see its affects on more types of media. Again, the main difference here is in the process of the creation of the bit-string. We have taken mp3 files and read the 16 bit amplitudes associated with them. There were two vectors corresponding to left and right. Then we converted these numbers to binary (as here the numbers could be negative as well we used the 2s complement notation to convert them to binary) and concatenated them to get two separate bit strings. Entropy expansion was applied to both the bit strings and another mp3 file was created from these. \\
The original file contained music and it was observed that the transformed file contained static noise. The ENT test can also be applied to this case, although it has to be noted that the binary data of music files contains other information apart from the amplitude information and since we are only modifying that part, this test will not be a good measure of entropy expansion. Still, parameters like the serial correlation coefficient showed an improvement from a value of 0.083505 for the original music file to a value of 0.056021 for the modified file. 
\section{Conclusions}\label{Conclusion}
Randomness is an essential feature of the cipher data. Deterministic PRNGs and low entropic TRNGs create a situation of entropy starvation, thus exposing data for cyber attacks. Resource constrained IoT devices are more prone to such attacks because of their low memory and processing power. In this work, we address the problem of entropy starvation by proposing an entropy expansion algorithm using n-qubit permutation matrices.

We have expanded the entropy of English text, images and audio files. Since videos are just concatenation of several images and audios, we can also apply entropy expansion to video files. Our algorithm works for all these file types without altering the file size. In particular, we have expanded the entropy by 75$\%$ in the worst case considered. We have observed that the entropy expansion increases in both scenarios: 1. When we increase the dimension of the permutation matrix. 2. When we increase the number of permutation matrices in the set. However, the expansion is more when we increase the dimension of the permutation matrix. This is expected because information space increases exponentially as we increase the size of the matrix. Same behaviour is observed when we modified the images using our protocol (Fig. \ref{scene}). One can see that the modified image does not have any observable pattern as we increase the permutation matrix to 15-qubit (Fig. \ref{blue_bird}). 

There are several offshoots of the present work. We have demonstrated entropy expansion using permutation matrices which modify the binary information of the qubits. One can study the entropy expansion capacity of the generalized permutation matrices (taking non-zero scalar as $1,2, \cdots, d$) by modifying the information of qudits. We anticipate further entropy expansion because of even larger information space of the generalized permutation matrices. Another potential application is to expand the entropy of streaming data for practical usage. FPGA implementations of our protocol will further increase the entropy expansion speed.

\section*{Acknowledgement} 
SG acknowledges the SPARK grant of IDex Open challenge 2.0 for the financial support.

\appendix
\onecolumngrid

\end{document}